\documentclass[aps,prb,twocolumn,floatfix,10pt,showpacs,superscriptaddress]{revtex4-1}
\usepackage{amsmath,amssymb,mathtools}
\usepackage{graphicx}
\usepackage[ansinew]{inputenc}
\usepackage{xcolor}
\usepackage{hyperref}

\newcommand\siesta{\textsc{Siesta}}
\newcommand\tsiesta{\textsc{TranSiesta}}

\begin{document}

\title{Current-induced atomic forces in gated graphene nanoconstrictions}
\author{S. Leitherer}
\affiliation{Center for Nanostructured Graphene, Department of Physics, Technical University of Denmark, DK-2800 Kongens Lyngby, Denmark}
\author{N. Papior}
\affiliation{Department of Applied Mathematics and Computer Science, Technical University of Denmark, DK-2800 Kongens Lyngby, Denmark}
\author{ M. Brandbyge}
\affiliation{Center for Nanostructured Graphene, Department of Physics, Technical University of Denmark, DK-2800 Kongens Lyngby, Denmark}

\begin{abstract}
Electronic current densities can reach extreme values in highly conducting nanostructures where constrictions limit current.
 For bias voltages on the 1 volt scale, the highly non-equilibrium situation can influence the electronic density between atoms, leading to significant inter-atomic forces. An easy interpretation of the non-equilibrium forces is currently not available. In this work, we present an \textit{ab-initio} study based on density functional theory of bias-induced atomic forces in gated graphene nanoconstrictions consisting of junctions between graphene electrodes and graphene nano-ribbons in the presence of current. We find that current-induced bond-forces and bond-charges are correlated, while bond-forces are not simply correlated to bond-currents.
We discuss, in particular, how the forces are related to induced charges and the electrostatic potential profile (voltage drop) across the junctions. For long current-carrying junctions we may separate the junction into a part with a voltage drop, and a part without voltage drop. The latter situation can be compared to a nano-ribbon in the presence of current using an ideal ballistic velocity-dependent occupation function. This shows how the combination of voltage drop and current give rise to the strongest current-induced forces in nanostructures.
\end{abstract}

\maketitle
\clearpage

\section{Introduction}

The current densities in nano-scale, ballistic conductors can reach extreme values compared to macroscopic Ohmic conductors.
For example, the break-down voltages of atomic chains of Au are beyond 
$1\,\mathrm V$ corresponding to a current-density\cite{Sabater2015} on the order of $10^{10}\,\mathrm{A}/\mathrm{cm}^2$, and the current-carrying capacity of narrow graphene conductors
can reach almost $10^9\,\mathrm{A}/\mathrm{cm}^2$ before breakdown.\cite{Moser2007a} 
From a technological point of view the nano regime poses challenges in terms of stability and reproducibility,
since in this extreme scaling limit the position of a few atoms control the device operation.

On the other hand, atomic control of the structure by external driving forces offers an enormous potential for further downscaling.
F.ex. it has been demonstrated in experiments how the current/field may be used to toggle switch atomic-scale contacts between different conductance states 
corresponding to different atomic configurations of metallic nano-contacts.\cite{scheer2013} 

In this paper we will concentrate on another important example, namely graphene nanostructures, which are now being created and changed using high applied voltages
and consequently electrical current.
Due to its excellent electrical and mechanical properties, graphene is a promising material for two-dimensional (2D) nanoelectronic applications.\cite{Geim07}
So-called ''electro-burning`` has been employed in experiments to fabricate nano-gaps between graphene electrodes.\cite{Sadeghi2658} 
These electrodes of single or few-layer graphene has in some cases subsequently been bridged by single molecules.\cite{Prins2011,weber2015,C7NR00170C,Sun2018}
Using similar techniques, the fabrication of electrically switchable graphene break junctions has been reported.\cite{zhang2012,standley2008}
Electron microscopy allows for structural, atomic-scale studies of graphene structures in the presence of high current and applied voltage.\cite{Harris2017} 
It has been seen how the structure of edges are changed by the current/voltage\cite{Jia09} or how layers fuse.\cite{Barreiro2012}
Current-induced motion/cleaning of adsorbed species on graphene has also been investigated.\cite{Moser2007a}

Under high bias and current density a number of different, possibly intertwined, effects play crucial roles
for the atomic configuration, such as motion driven by locally induced fields, Joule heating and temperature gradients,
as well as current-induced forces due to a steady momentum transfer from electronic current to ions.\cite{Dundas2009CurrentdrivenAW,DiVentra2002,JT2015}
Common for these structures and effects in graphene nanostructures is that the electrons may to a large degree be in the ballistic quantum transport regime,
as seen e.g. by the appearance of interference phenomena.\cite{Sadeghi2658,Garcia-Suarez2018,Gehring2016}
Experiments performed at high voltage bias on a bilayer constriction show an uniaxial lattice expansion of more than 5\% at a current density
on the order of $10^9\,\mathrm{A}/\mathrm{cm}^2$ before breaking.\cite{Borrnert2012}

The understanding of the role of voltage and currents in such systems and processes are still rudimentary.
We consider here a simple, narrow graphene ribbon system using first principles calculations based on Density Functional Theory
combined with Non-Equilibrium Greens functions (DFT-NEGF). 
We have previously studied the electron-phonon interaction in transport and the voltage drop dependence on gating in this system.\cite{tue2016,nick2016} In this paper we consider the current-induced forces in the presence of steady-state electronic current, and analyze these in terms of
the changes in electronic distributions. 

Our DFT-NEGF calculations, presented in the first part of this work, return forces for systems that are defected in the sense of having a scattering region. 
However, for ballistic bulk systems one could
imagine a current flowing which is far from any scattering potential, 
and approximately behaves as though states are occupied depending on their velocity. 
We present this approach and compare to the DFT-NEGF forces in the last part of this paper
(Sec.~\ref{sec:bulk}).

\section{Setup and Method}
\label{sec:method}

\begin{figure}[t]
 \includegraphics[width=0.45\textwidth]{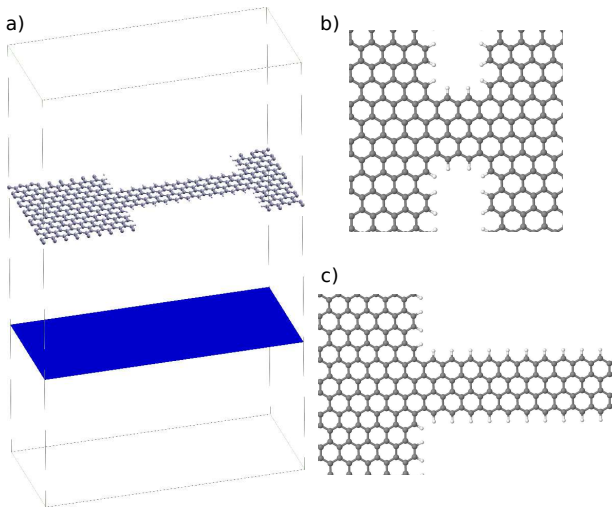}
  \caption{Graphene nanoconstrictions with electrostatic bottom gate. a) represents the unit cell of a generic constriction, with the blue square indicating the position and shape
  of the electrostatic gate. Edge atoms of the graphene are saturated by hydrogen in order to avoid dangling bonds. b) Short constriction and c) semi-infinite GNR constrictions considered in this work. }
  \label{fig:001}
\end{figure}

The systems we investigate are constrictions consisting of graphene nanoribbons (GNRs) of varying lengths, 
placed between graphene electrodes (cf.~Fig.~\ref{fig:001}).
In addition, the junctions are electrostatically gated.
Their geometries are relaxed at zero bias using the \siesta\ package and their properties are studied at finite bias
using the nonequilibrium electronic transport package \tsiesta . Computational details are described in Sec. \ref{sec:comp}.

We apply the field-effect gate model of Ref.~\onlinecite{nick2016}. 
A charged plane is placed at $15\,\mathrm{\AA}$ underneath the graphene constriction. 
The plane carries a charge density of $ n= g\cdot10^{13} e^-/\mathrm{cm}^2$, where $g$ defines the gating levels,
with $g<0$/$g>0$ referring to $n$/$p$-doping.
Placing the electrostatic gate allows for a tuning of the conductance of the junction, while  
on the other hand, the position of the voltage drop in the constriction can be controlled.\cite{nick2016}
Thus we explicitly include the role of the gate-induced carriers on the screening properties and potential profile.

We focus on two distinct geometries, shown in Fig.~\ref{fig:001} (b,c):
The first (b) is a graphene constriction with a very short GNR, and
the second junction (c), consists of a large region of pristine graphene connected to a semi-infinite GNR.
The results are presented in Sec.~\ref{sec:gnr1}, Sec.~\ref{sec:gnr2}, respectively.

Since our aim here is to study generic features of the local current and potential drop, and the relation to inter-atomic forces, we neglect the role of spin-polarization at the zig-zag edges.\cite{Magda2014}

\begin{figure}[b]
 \includegraphics[width=0.45\textwidth]{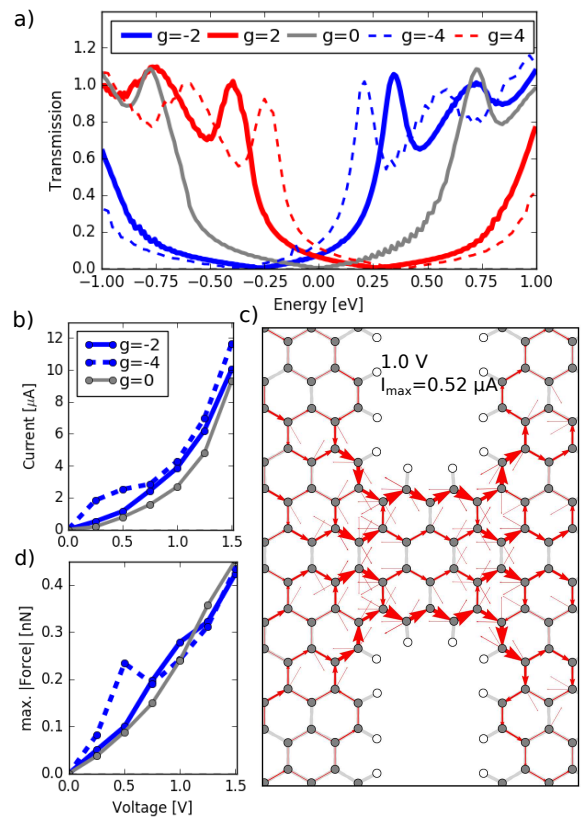}
  \caption{a) Zero-bias transmission for different gate charges $g$ (gray without gate), b) total current as a function of bias voltage for selected gate charges, 
  c) real-space bond currents at $1\,\mathrm V$, $g=-2$, 
  and d) maximal absolute force over bias voltage for different gate charges. 
  }
  \label{fig:002}
\end{figure}

\section{Results and Discussion}

\subsection{Short constriction}
\label{sec:gnr1}

We first consider the left-right symmetric graphene nanojunction with a short GNR (cf.~Fig.~\ref{fig:001} (b)). We employ periodic boundary conditions in the direction transverse to the constriction with a corresponding $k$-point sampling.

\subsubsection{Conductance properties}

In  Fig.~\ref{fig:002} (a-c), we discuss the transport properties of the junction, 
in particular the transmission probability, total currents, and real-space ''bond currents``. In the following we implicitly assume $k$-dependence.
Thereby, the transmission through the constriction is calculated from,
\begin{equation}
 T(E,V)=\operatorname{Tr}\left[ \mathbf G \boldsymbol\Gamma_L \mathbf G^{\dagger} \boldsymbol\Gamma_R \right],
\end{equation}
where $\mathbf G = [(E+i\eta)\mathbf S - \mathbf H-\boldsymbol\Sigma]^{-1}$ is the nonequilibrium retarded Green's function with device Hamiltonian $\mathbf H$, 
overlap $\mathbf S$ and selfenergies $\boldsymbol\Sigma=\sum_{\alpha=L,R}\boldsymbol\Sigma^{\alpha}$, 
and $\boldsymbol\Gamma^{\alpha}=i (\boldsymbol\Sigma^{\alpha \dagger} - \boldsymbol\Sigma^{\alpha})$.
The total current is given by,
\begin{equation}
 I(V) = \frac{2e}{h}  \int T(E,V) [f_L(E,V)- f_R(E,V)]\, \mathrm{d}E,
\end{equation}
with $f_{L/R}$ being the Fermi distributions in the electrodes, 
where the chemical potentials at finite bias are shifted according to $\mu_{L/R}= E_F\pm eV/2$.

Fig.~\ref{fig:002}a shows the zero-bias transmission for different values of the gating parameter $g$. 
Gating leads to a doping of the junction, i.e.~the charge-neutrality point in the DOS is shifted relative to its position at $g=0$. 
Accordingly the transmission is shifted further into the conductance window with $g$.
This results in a higher conductivity especially at small bias, cf.  total current in Fig.~\ref{fig:002}b.
In the high bias regime the current is not significantly enhanced by the gating, because the transmission at high energies is nearly $1$.

\begin{figure}[t]
 \centering
 \includegraphics[width=0.48\textwidth]{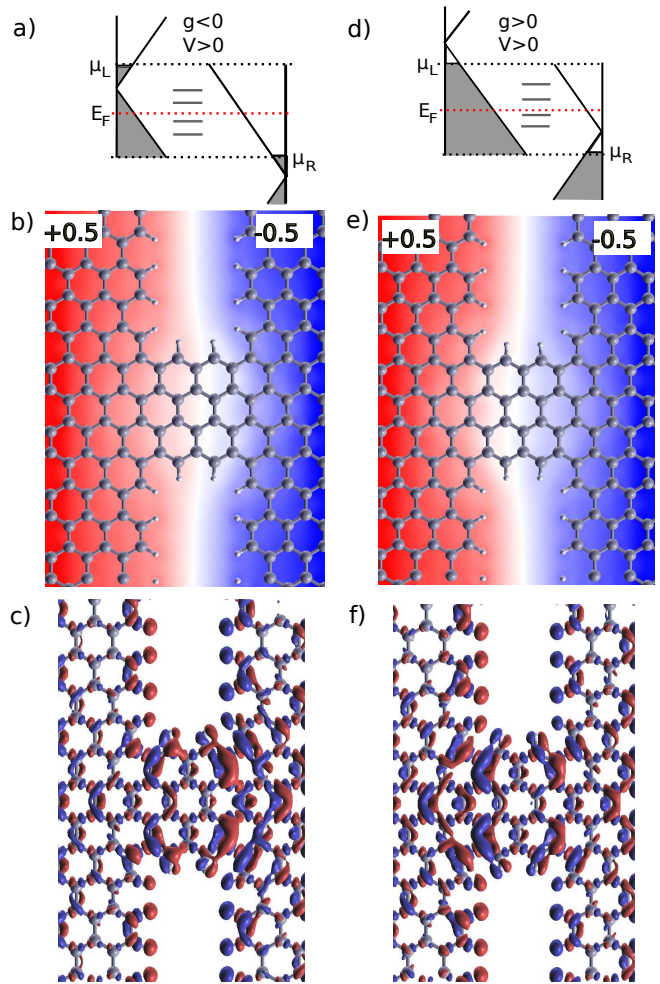}
  \caption{a) Energy scheme of the electrode DOS for negative $g$ and positive bias voltage, b) electrostatic potential profile, and c) bias-induced charge redistribution 
  at $1\,\mathrm{V}$ for $g=-2$. (d,e,f) DOS, potential and charges at $1\,\mathrm{V}$ for $g=+2$. }
  \label{fig:003}
\end{figure}

A spatial distribution of the current flowing through the junction can be obtained by calculating bond currents.\cite{Nakanishi2001,todorov2002,solomon2010}
The energy-dependent spectral bond currents from atom $n$ to $m$ are defined as,
\begin{equation}
 \partial J_{nm}(E,V) = \frac{2e}{\hbar} \sum_{\mu,\nu} \operatorname{Im} \left\{ \mathbf A^{\alpha}_{\mu\nu}(E,V) [\mathbf H(V) - E\mathbf S]_{\nu\mu}\right\}
 \label{eqn:curr}
\end{equation}
where $\alpha=L,R$ refers to the electrode, $\nu \in n$ and $\mu\in m$ are orbital indices, and 
the spectral function is given by
\begin{equation}
  \mathbf A^{\alpha}(E,V) = \mathbf G\boldsymbol \Gamma^\alpha\mathbf G^\dagger.
\end{equation}
The bond current is obtained by integrating Eq.~\eqref{eqn:curr} over the Fermi window, defined by $f_L-f_R$ at the corresponding bias:
\begin{equation}
  J_{nm}(V) =  \frac{1}{2\pi}\int \partial J_{nm}(E,V) [f_L(E,V)- f_R(E,V)]\, \mathrm{d}E.
 \label{eqn:curr1}
\end{equation}
In Fig.~\ref{fig:002}c the bond currents at a bias of $1\,\mathrm V$ are shown.
The highest current density appears at the entrance to the constriction and along the edge atoms in the constriction. 
In the pristine graphene, bond currents obtain smaller values and spread out across the lattice.
They obey the law of particle conservation, i.e.~through any section dividing the left and right part their total sum is conserved.
The current pattern exhibits a somewhat left-right/top-bottom symmetry, which obviously stems from the junction symmetry. 

\begin{figure*}[t!]
 \centering
 \includegraphics[width=0.95\textwidth]{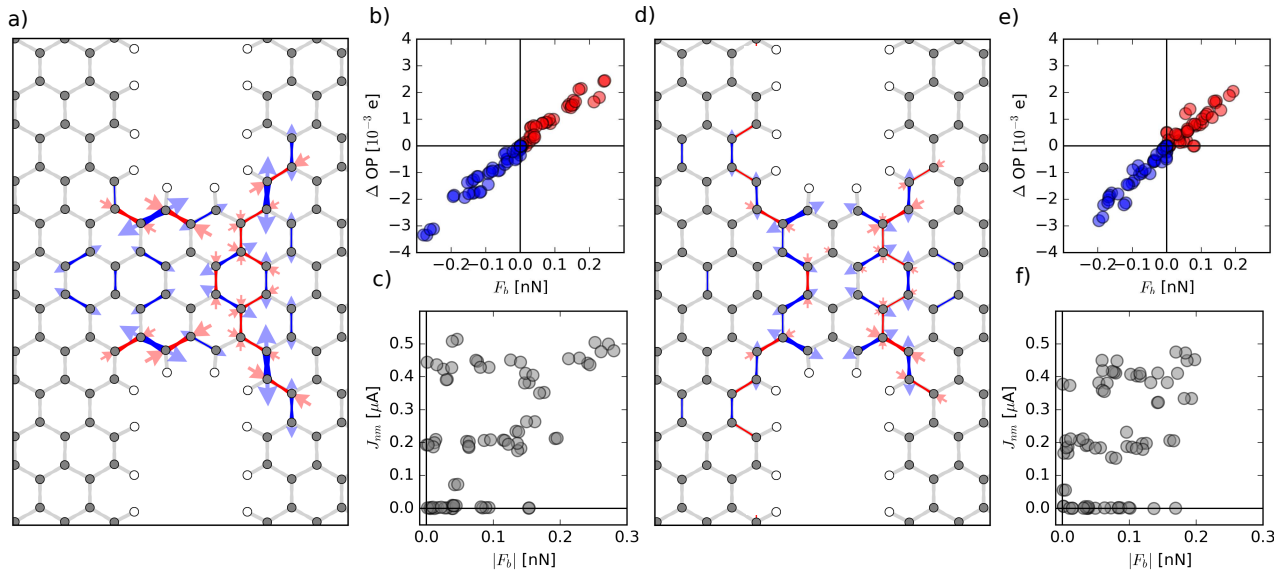}
  \caption{a) Induced bond forces, $F_b$, and change in overlap population $\Delta$OP at 1.0 V, $g=-2$. The forces $F_b$ are shown as vectors in light red and blue.
  Force vectors are pointing inwards (outwards) to indicate bond compressing (stretching), while the vector thickness corresponds to the force strength.
  The change in OP is depicted as in-/decreasing density along the bond, in red for positive and blue for negative $\Delta$OP. 
  Induced forces/charges below a cutoff of $|F_b|=0.004$ nN/$\Delta$OP$=1\cdot 10^{-4}$ e are set to zero.
  b) Correlation between bond forces $F_b$ and $\Delta$OP, c) correlation between bond forces $|F_b|$  and bond currents $J_{nm}$,
  (d,e,f) equivalent pictures for $g=+2$.}
  \label{fig:004}
\end{figure*}

We want to study the interatomic forces in the graphene constriction, which are induced when a finite bias voltage is applied.
These forces are calculated from the non-equilibrium electron density defined by the density matrix $\mathbf D$, which we obtain 
from the DFT-NEGF formalism.\cite{PhysRevB.65.165401}
In particular, the force acting on atom $n$ with coordinate $\vec R_n$ is given through the force operator $\vec{\mathbf{F}}_n$ and the  density operator $\mathbf D$ via
\begin{equation}
 \vec F_n =  \operatorname{Tr}\left[\vec{\mathbf{F}}_n \mathbf D\right] =-\operatorname{Tr}\left[\frac{\partial \mathbf H}{\partial \vec R_n} \mathbf D\right],
 \label{eqn:force}
\end{equation}
and the non-equilibrium density operator,
\begin{equation}
 \mathbf D = \int \left[\mathbf{A}^L(E,V)f_L(E,V) -\mathbf{A}^R(E,V)f_R (E,V)\right]\mathrm{d}E.
\end{equation}

In Fig.~\ref{fig:002}d, we plot the maximum absolute force induced by the non-equilibrium between all atoms in the short GNR constriction,
depending on the gate parameter and the bias voltage (for a spatial distribution of the forces, see below).
The maximum force is seen to increase with voltage roughly following the current, where as both are more weakly influenced by the gate parameter.
We find forces of $\sim0.2\,\mathrm{nN}$ at $1\,\mathrm{V}$.

Theoretical models were compared to tunnel-to-contact experiments of atomic point contacts in order to explicitly relate the conductance to the atomic forces at low bias $\sim 1$mV, as for example presented in Ref.~\onlinecite{Ternes2011}.
Below, we will present a detailed analysis of the forces and compare these to the local current and potential drop at the higher voltages.

\subsubsection{Potential drop and finite bias charge redistribution}

At finite bias, the chemical potential in the electrodes is symmetrically shifted and 
an electrical field between the electrodes exists across the junction, resulting in a rearrangement of charge. 
We present in Fig.~\ref{fig:003}a the schematic picture of the electrode density of states (DOS) and the energy levels of the junction at finite bias for $g=-2$, 
(b) the electrostatic potential landscape $\Phi(1\,\mathrm{V},g=-2) - \Phi(0\,\mathrm{V},g=-2)$ and (c) the induced charge
$\rho(1\,\mathrm V,-2) - \rho(0\,\mathrm V,-2)$ in the short GNR constriction.
In Figs.~d-f we present the same analysis for $g=+2$.

The energy scheme, a) and d), illustrates how the non-symmetric coupling is induced via the electrode having the largest DOS in the bias window. This results in an electrostatic potential pinning of the contriction. In b) it pins to the right electrode presenting the larger DOS in the voltage window, and opposite for the case in e), see Ref.~\onlinecite{nick2016} for details. Such relative changes in the electrostatic potential also results in a different charge redistribution. In c) and f) we show how the charge redistribution is highest at the interface of the potential drop.

Our analysis shows that forces are highly correlated with such charge redistributions and in the following we will outline simple relations between the charge redistributions and forces.

\subsubsection{Bond forces and overlap population}

To simplify the representation of the forces, Eq.~\eqref{eqn:force}, we project them onto the atomic bonds.
These bond forces are defined as the difference of the forces on atom $n$ and $m$,  projected onto the bond vector ${\vec r}_{nm} =  \vec r_m - \vec r_n$:
\begin{equation}
 \vec F_{b,nm} = \frac{ (\vec F_m - \vec F_n ) \cdot {\vec r}_{nm}}{|{\vec r}_{nm} | }
  \label{eqn:bforce}
\end{equation}
With this definition, positive (negative) bond forces can be interpreted as compressive (repulsive). 
Note that our structures are relaxed at zero bias, thus $F_b$ refers to the bias induced forces.

Fig.~\ref{fig:003}a depicts induced bond forces at $1\,\mathrm V$ for $g=2$ (d for $g=-2$).
Compressive (repulsive) bond forces are shown as arrows in light red (light blue).
We draw the force arrows at both atoms of a bond to indicate if the force is stretching or compressing the bond. 

In a),d) we also show how we can relate the forces to the charge redistribution in the junction. 
In particular, we have calculated the amount of charge in the bonds, 
also termed overlap population (OP), similar to the analysis in Ref.~\onlinecite{PhysRevB.67.193104}. 
This approach is based on interpreting the bond population as a measure of the bond strength. \cite{mull}
The OP is given by a sum over atomic orbitals $(i,j)$ belonging to the atoms $n,m$,
 \begin{align}
  \mathrm{OP}&= \sum_{\alpha=L,R}\sum_{\substack{i\in n \\ j\in m}} \mathbf O_{ij}^{\alpha},
  \shortintertext{with}
  \mathbf O^{\alpha}_{ij} &= \mathbf S_{ij}\int \mathrm{d}\epsilon\, \mathbf A^{\alpha}_{ij}(\epsilon) f_{\alpha}(\epsilon-\mu_{\alpha}).
  \label{eqn:002}
 \end{align}
and the spectral function $\mathbf A^\alpha$, where $\alpha=L,R$, since it has contributions from left and right-originating states.
To obtain the bias-induced bond charge, we calculate the change in overlap population with bias, $\Delta\mathrm{OP} = \mathrm{OP}(V)-\mathrm{OP}(0)$.

In Fig.~\ref{fig:003}a and d, the nonzero $\Delta\mathrm{OP}$ are depicted as density along the bond; in particular the line thickness corresponds to $|\Delta\mathrm{OP}|$ and red (blue) indicates if it has positive (negative) sign.  
We find that the bond forces and the change in overlap population are clearly correlated. An increase (decrease) of charge in the bond corresponds to a positive (negative) bond force,
corresponding to bond elongation (compression).
This correlation between bond force and population is also revealed by the scatter plots in  Fig.~\ref{fig:003}b for $g=2$ (and e for $g=-2$).

Similarly to the bond forces, we plot the bond currents in Fig.~\ref{fig:002}c. While the bond-currents show a left-right symmetry, this symmetry is fully absent in the bond forces. As shown in Fig.~\ref{fig:003}c and f, the bond current and force strength do not clearly correlate. This suggests that even though certain atoms experience a high current density it is not necessarily reflected in forces acting on it, or at least its effect is minor compared to other effects.
We note that recent work calculating the current-induced forces in graphene nanoribbons based on single-orbital tight-binding model find a correlation between the local currents and bond-forces.\cite{Asoudegi2019} But this clearly will depend on the level of description of the connection to electrodes and the associated potential drop.

\begin{figure}
 \centering
\includegraphics[width=0.5\textwidth]{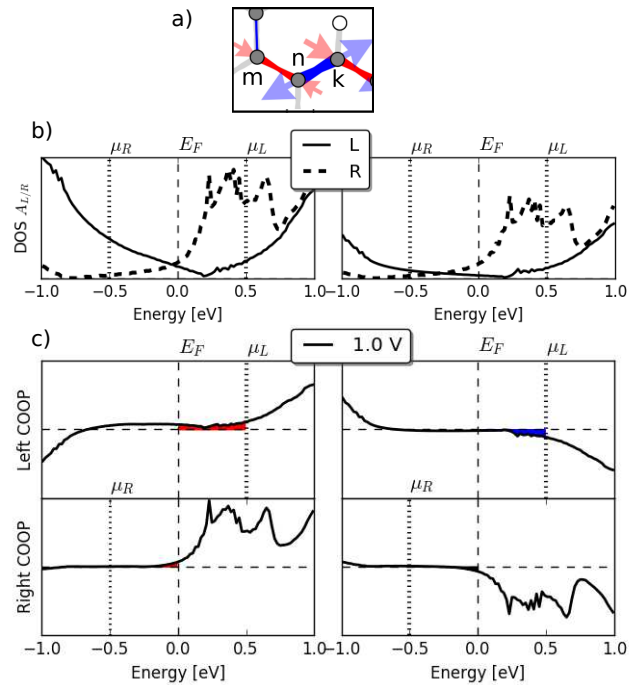}
            \caption{ a) Chemical bond of atom $m$ and $n$ with compressive bond force (red), and bond of atom $n$ and $k$ with repulsive bond force (blue) at $1\,\mathrm V$ and $g=-2$. 
            b) DOS of left-/right-going states, $\mathbf A^L$ and $\mathbf A^R$, on atoms $m$ and $n$ (left panel) and atoms $n$ and $k$ (right panel) at $1\,\mathrm V$.  
                States in $\mathbf A^L$ ($\mathbf A^R$) below $\mu_L=0.5$ eV ($\mu_R=-0.5\,\mathrm{eV}$) are occupied. 
            c) COOP analysis for bond $m$-$n$ (left panels) and bond $n$-$k$ (right panels).  
            Bond $m$-$n$ is strengthened by current, as at $1\,\mathrm V$ bonding states get populated in  $\mathbf A^L$ (red shaded area in left COOP). The DOS that is depleted in $\mathbf A_R$ is small.
            Bond $n$-$k$ is weakened by current, as antibonding states get filled (blue shaded area).
            }
            \label{fig:005}
\end{figure}

To get further insight into the bias-induced bond populations, we analyze the crystal orbital overlap population (COOP) curve \cite{RevModPhys.60.601},
which is the energy-resolved overlap population.
For the bond between atoms $n$ and $m$, the COOP is defined as
\begin{equation}
\mathrm{COOP}(E) = 2 \sum_{i\in n, j \in m} \mathbf S_{ij}\mathbf A_{ij}^{\alpha}(E), 
\end{equation}
with $\alpha = L,R$ referring to left- and right-coming states.
The sign of the COOP curve determines whether the states contributing to the bond have bonding (positive) or anti-bonding (negative) character.\cite{RevModPhys.60.601} 
Therefore filling of bonding/depletion of antibonding states will lead to a strengthening of the bond force, and vice versa.
Note that integrating the COOP (weighted by the Fermi distribution) gives the OP.

In Fig.~\ref{fig:005} we present this analysis for two bonds in the junction of Fig.~\ref{fig:003}a, which experience a high bond force: 
One bond that is compressed, $m$-$n$, and one that is stretched, $n$-$k$, under influence of the current (cf.\ Fig.~\ref{fig:005}a). 
The DOS of left- and right traveling states at $1\,\mathrm{V}$, shown in b), is similar for atoms $m$-$n$ (left panel) and atoms $n$-$k$ (right panel). 
However, for atoms $m$-$n$, the states that are energetically located in the conductance window have bonding character, as indicated by a positive COOP (left panels in c), 
while on atoms $n$-$k$ they are antibonding (right panels in c). 
Moslty relevant are states in $\mathbf A^L$, since those get filled by shifting $\mu_L$ up, while the occupation of right states (bottom panels) does not change significantly by the downshifting of $\mu_R$.
The positive $\Delta\mathrm{OP}$ for the bond $m$-$n$ can be traced back to an increased filling of bonding states, 
while on atoms $n$-$k$ antibonding states become occupied, resulting in a negative $\Delta\mathrm{OP}$.

\begin{figure}[b]
 \centering
 \includegraphics[width=0.5\textwidth]{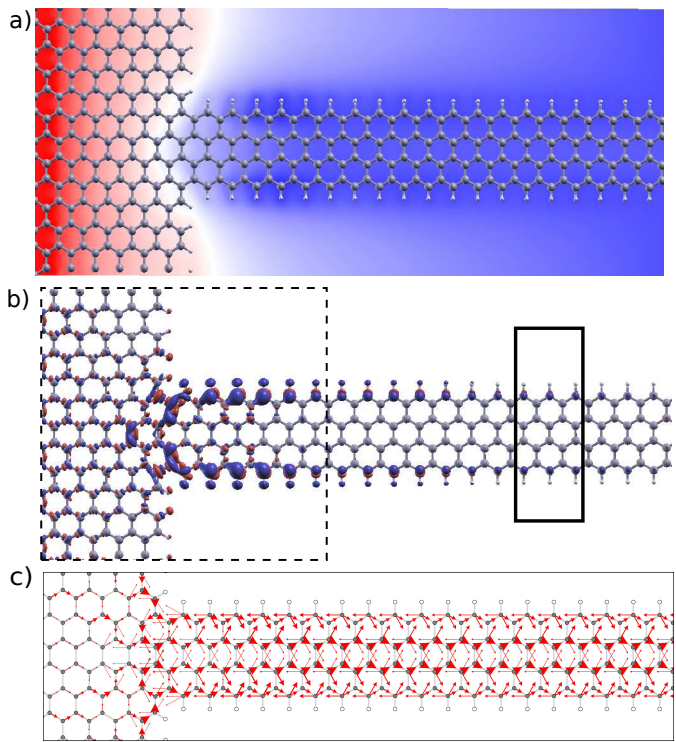}
  \caption{ a) Electrostatic potential profile b) bias-induced charge redistribution and c) bond currents in the wide-narrow GNR constriction at 0.75 V, $g=-2$. }
  \label{fig:006}
\end{figure}
\begin{figure*}[]
 \includegraphics[width=0.95\textwidth]{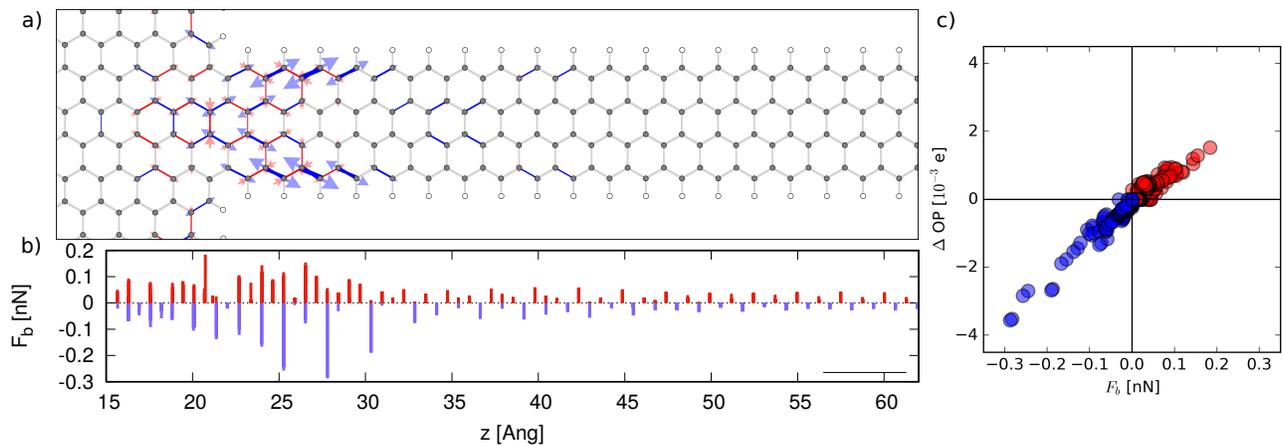}
  \caption{ a) Distribution of induced bond forces $F_b$ and $\Delta$OP in the GNR constriction, b) maximal values of $F_b$ along the transport direction $z$, 
  c) correlation between $F_b$ and $\Delta$OP, at 0.75 V, $g=-2$.  }
  \label{fig:007}
\end{figure*}

\subsection{Wide-narrow GNR constriction}
\label{sec:gnr2}

In order to study in more detail the influence of the potential profile on the forces, we consider a wide-narrow constriction, 
where the right electrode is a semi-infinite GNR (cf.~Fig.~\ref{fig:001}c).  
We focus on a negative doping of $g=-2$ and positive bias voltages.

\subsubsection{Potential drop, charge redistribution, bond currents and forces at non-equilibrium}

In longer GNR constrictions, a pinning of the potential to one of the electrodes can be achieved, 
leading to a very localized potential drop at the transition between the GNR and the graphene.
In Fig.~\ref{fig:006}a we show the potential profile at positive bias (here $0.75\,\mathrm V$) in a graphene constriction with a horizontally extended GNR.
For a detailed discussion of the bias-dependence of the voltage drop in very similar GNR constrictions, we refer to Ref.~\onlinecite{nick2016}.

In Fig.~\ref{fig:006}b, we show the the bias-induced charge density.
We find that the largest amount of charge is accumulated near the potential drop.
This is due to the fact that at finite bias, reflection of incoming channels takes place at the scatterer, i.e.~the constriction entrance where the potential drop is located.
These scattering processes induce Landauer dipoles in this region.\cite{Landauer}
Farther away from the potential drop, we find a smaller amount of induced charge density in the GNR, which converges at longer distances.
As the states closest to $E_F$ are edge states, this charge is  mainly localized on the zigzag edges of the GNR.

Figure~\ref{fig:006}c depicts bond currents through the extended GNR constriction. Due to particle conservation, the bond currents are of the same size all along the GNR. Note that there is little to no correspondence between the charge redistribution and bond currents. In a similar way as the charge density profile, the non-equilibrium forces are maximal in the region of the potential drop. 
This is illustrated in Fig.~\ref{fig:007}a, where we show the change in bond forces and overlap population. 
Again, we find compressive/repulsive forces for bonds where a large amount of bond charge is induced/depleted (cf.~Fig.~\ref{fig:007}c).
The maximum/minimum forces along the transport direction in the junction are depicted in Fig.~\ref{fig:007}b.

For a detailed analysis of the forces, 
two regions can be distinguished in the junction:
One is the region of the wide-narrow transition (dotted square in Fig.~\ref{fig:006}b), where the potential drop is located.
Here we find the largest forces with maximum strength of $|F_b|\le0.38\,\mathrm{nN}$. 
In this region there are contributions to the forces from the reflected charge density as well as from the density of transmitting channels, beyond the usual electrostatic forces.

Deeper in the GNR, the forces and bond populations become significantly smaller ($|F_b|\le 0.05\,\mathrm{nN}$) and reach a periodic pattern. 
In this region (bold square in Fig.~\ref{fig:006}b), the electrostatic potential profile is very flat and nearly equivalent to the right electrode chemical potential, $\mu_R$.
Thus, the forces in this region are not related to a potential drop, but are ideally solely originating from the flow of current.
The correspondence between induced bond forces and $\Delta\mathrm{OP}$ is still given,
with the accumulated charge coming from the current in the occupied, transmitted conductance channels.

\subsubsection{Forces without voltage drop}
\label{sec:bulk}

In the extended GNR constriction, we have studied forces in region 2, where a flat potential profile has established. 
This allows for a comparison with a perfectly ballistic bulk system, where
a current flows without the electrical field in the potential drop. This enables the use of periodic boundary conditions and a Bloch band description. 
Specifically, we may employ a bulk-like calculation scheme where states are occupied according to their band velocity,
\begin{equation}
{v}_{nk}=\frac{1}{\hbar}\frac{\partial\varepsilon_n(k)}{\partial k}
\end{equation}
where $n$ is the band-index.
The idealized, ballistic occupation function corresponds to a situation where current is fed into the nano-ribbon from ideal electrodes without any scattering in the voltage window. 
This is of course idealized and will overestimate the current.
The non-equilibrium distribution function relative to equilibrium is,
\begin{equation}
\delta f(nk)=\Theta(v_{nk}\cdot \hat{\mathbf e})
\left[f_L(\varepsilon_n(k),V)-f_R(\varepsilon_n(k),V)\right]
\end{equation}
where $\Theta$ is the Heaviside step function and
$\hat{\mathbf e}$ is the direction of the external bias driving the current. The chemical potentials for left- and right-movers are $\mu_L=E_F+eV/2$ and $\mu_R=E_F-eV/2$ with $V$ being the applied voltage. The quasi-Fermi level, $E_F$, is determined in the self-consistent DFT cycle such that the charge is neutral in the unit-cell. We will denote this type of calculations as {\em ballistic-bulk} calculations. 

\begin{figure}
 \centering
\includegraphics[width=0.47\textwidth]{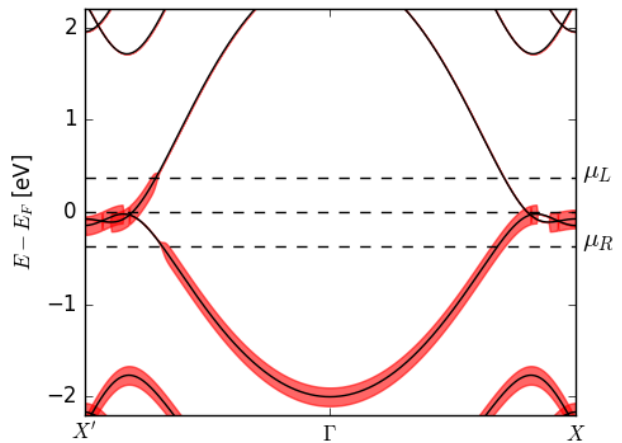}
  \caption{ Bulk-bias applied to GNR. 
 Shown is the bandstructure, where a positive bias of 0.75 V along the GNR direction changes the
occupation of right-moving states with positive velocity such that states are filled up to $\mu_L = eV/2$. Similarly are the
left moving states with negative velocity emptied above $\mu_R = -eV/2$. Filled bands are indicated in red.  }
  \label{fig:008}
\end{figure}

We have performed a ballistic-bulk calculation for the GNR for $g=-2$. Fig.~\ref{fig:008} illustrates the filling of bands in the GNR at a bulk bias of 0.75.

\begin{figure}[t]
 \centering
\includegraphics[width=0.45\textwidth]{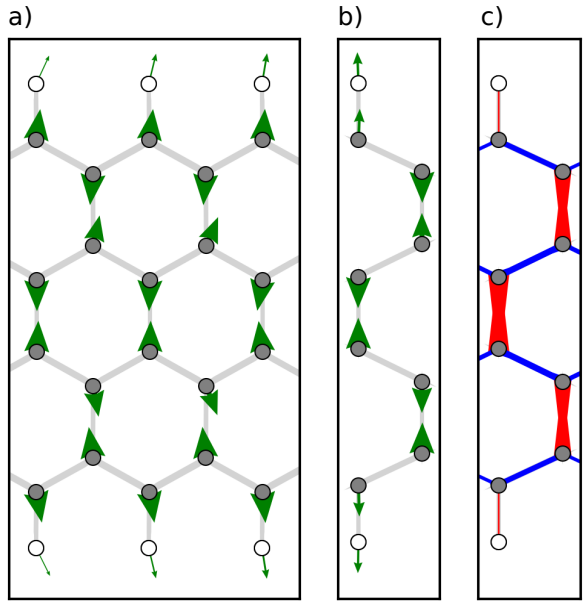}
  \caption{ a) Force pattern far away from the potential drop at $0.75\,\mathrm{V}$ from \tsiesta\ calculation, b) forces and c) $\Delta\mathrm{OP}$ from bulk calculation at $0.75\,\mathrm{V}$.  }
  \label{fig:009}
\end{figure}

In Fig.~\ref{fig:009},
we compare the DFT-NEGF forces in the constriction far away from the potential drop
with the forces from the ballistic-bulk calculation. 
The cutout in Fig. \ref{fig:009}a corresponds to the bold square in Fig.~\ref{fig:006}b, while \ref{fig:008}b shows the unit cell of the ballistic-bulk calculation. 
We recover a very similar force pattern for both calculations, with forces only perpendicular to the transport direction. 
In both setups, the inner atoms of the GNR are contracting,
while the edge atoms and hydrogen are slightly pushed outwards.
The correlation between bond forces and induced bond populations is also revealed in the ballistic-bulk calculation, cf.~Fig.~\ref{fig:009}c.
Due to the symmetry of the single unit cell in the bulk calculation, it returns symmetric forces.
The DFT forces show some deviations, as Fig.~\ref{fig:009}a is taken out from the large junction.
Also, they are lower in magnitude compared to the bulk forces. The maximum bulk forces in Fig.~\ref{fig:009}b are $0.2\,\mathrm{nN}$ for $0.75\,\mathrm{V}$, comparable to the forces in the region of the voltage drop in the constriction.  
On the other hand the DFT-NEGF forces in the ''bulk`` part of the constriction are $\sim 0.05\,\mathrm{nN}$, and thus a factor of 4 smaller than ballistic-bulk, while the ratio of the currents at this voltage is, however, roughly a factor of 20. So it is clear that the quasi-gap in the transmission seen in Fig.~\ref{fig:002}a due to the connection to the graphene electrode is important.

\section{Summary}

Summing up, we have analyzed non-equilibrium forces due to the presence of current in graphene nanoconstrictions by employing first principles transport calculations.
We have shown that the induced forces are related to a rearrangement of bond charges due to left/right incoming scattering states.
The forces and charges are maximal in the region where the potential drop takes place, because scattering happens there and dipoles are induced. 
We have further demonstrated forces which exist without potential drop and can thus be considered as purely ''current-induced``.  

Our theoretical work can be help to understand current-induced strains, bond-breaking processes \cite{Tod2001,Erp2018}, and mechanisms that lead to the destruction of devices at the atomic scale.

Funding by Villum Fonden (Grant No. 00013340) and the Danish Research Foundation
(Project DNRF103) for the Center for Nanostructured Graphene (CNG) is acknowledged.

\section{Computational details}

\label{sec:comp}

\subsection{DFT parameters}

The calculations were done using the \siesta/\tsiesta\ code with the PBE-GGA functional for exchange-correlation and a SZP basis-set.\cite{PAPIOR20178}
Spin polarization is not considered. The mesh cutoff was 300 Ry.
In \siesta\ we used an optimized k-point sampling according to the bias window. The transport calculations were averaged over 25 to 50 transverse k-points.
In the bulk calculations 1000 k-points along the ribbon are used.

Physical quantities like transmission, current, overlap population and COOP were extracted using TBtrans and SISL.\cite{zerothi_sisl}

\section{References}

\bibliographystyle{apsrev4-1}

%

\end{document}